\documentclass[conference]{IEEEtran}

\usepackage{amsmath,amssymb}
\usepackage{graphicx}
\usepackage{cite}
\usepackage{url}
\usepackage{siunitx}

\title{Scheduler-Agnostic Adaptive-FEC for MPQUIC: Field Evaluation over Commercial Cellular Paths}

\author{
\IEEEauthorblockN{
  Takuma Tsubaki,
  Soto Anno,
  Seiya Komatsu,
  Takashi Torii,
  Takuya Tojo}
\IEEEauthorblockA{
  Network Service Systems Labs., NTT, Inc., Japan\\
  Email: \{takuma.tsubaki,soto.anno,seiya.komatsu,takashi.torii,takuya.tojo\}@ntt.com}
}

\begin{document}
\maketitle

%======================
\begin{abstract}
  Multipath transport has long been studied to improve robustness over heterogeneous networks, and Multipath QUIC (MPQUIC) is particularly attractive for low-latency applications because QUIC Datagram enables transmission without retransmissions.
  However, this advantage comes with a key drawback: QUIC Datagram is vulnerable to packet loss, especially in vehicular cellular environments with handovers and rapidly varying radio conditions.
  While prior studies have explored adaptive-FEC and coded multipath designs, practical deployment remains challenging when such mechanisms require substantial modifications to the transport stack or scheduler.
  In this paper, we present an adaptive Forward Error Correction (FEC) scheme for MPQUIC Datagram that reuses QUIC loss detection signals and an off-the-shelf Reed--Solomon (RS) coding library.
  The sender estimates smoothed per-path loss rates from QUIC loss signals and adaptively determines the number of parity packets for each FEC block according to the expected packet loss.
  We implement the proposed scheme on an existing MPQUIC stack and evaluate it through vehicular field experiments over three commercial LTE/5G paths.
  Under the main 4.0\,Mbps setting, the proposed scheme improves both reliability and latency compared with No-FEC: the average one-way delay is reduced from 103.0\,ms to 70.8\,ms, the 95th-percentile delay from 281.2\,ms to 142.3\,ms, and the packet loss rate from 1.7\% to 0.8\%, with an average coding rate of 0.94.
  These results indicate that scheduler-agnostic adaptive-FEC can provide practical latency and reliability gains for MPQUIC Datagram under the measured vehicular mobile conditions.
\end{abstract}


%======================
\section{Introduction}

Remote monitoring is a key enabling function for autonomous vehicles, especially in Level 4 systems where vehicles operate without onboard drivers.
In such scenarios, sensor data, video streams, and control-related information must be transmitted in real time to maintain safe and stable operation.
However, achieving both low latency and high reliability over mobile networks remains difficult because vehicular cellular links are affected by handovers, fluctuating radio quality, and time-varying congestion~\cite{3gpp_ts_22186,denouden2022remote}.

Multipath transport has long been studied as a promising approach to improve communication robustness over heterogeneous and time-varying networks.
For example, Multipath TCP (MPTCP) enables a single connection to use multiple paths simultaneously and improves throughput and resilience against path degradation~\cite{rfc8684}.
However, for delay-sensitive applications such as remote monitoring, MPTCP still inherits TCP's retransmission-based recovery and in-order delivery semantics, which can increase latency through retransmissions and Head-of-Line (HoL) blocking~\cite{rfc8684}.

QUIC provides a different design point.
As a UDP-based transport protocol implemented in user space, QUIC offers low-latency extensibility and flexible evolution~\cite{rfc9000}.
In particular, the QUIC Datagram extension enables unreliable transmission without retransmissions, making it suitable for real-time applications that prefer timeliness over perfect delivery~\cite{rfc9221}.
Furthermore, Multipath QUIC (MPQUIC) extends QUIC to multiple paths and is therefore a promising platform for practical low-latency multipath communication in mobile environments~\cite{deconinck2017mpquic}.

This advantage, however, comes with an important drawback.
Because QUIC Datagram does not provide built-in loss recovery, application performance can degrade significantly under lossy conditions~\cite{rfc9221}.
This issue is particularly serious in vehicular cellular environments, where packet losses and delay spikes frequently occur due to mobility-related channel variations~\cite{denouden2022remote}.
Thus, although MPQUIC Datagram is attractive for low-latency communication, its robustness against packet loss remains a key challenge.

Forward Error Correction (FEC) is a natural approach because it enables packet recovery without retransmissions~\cite{rfc6363}.
Prior studies have investigated FEC for QUIC and coded multipath transport, including adaptive-FEC with an FEC-aware MPQUIC scheduler~\cite{vu2021supporting} and vehicle-to-cloud coded multipath streaming in real deployments~\cite{cellfusion2023}.
These studies indicate that combining multipath transport and coding is promising.

Nevertheless, an important practical question remains open for our target scenario.
Many coded multipath designs require tight integration with the transport stack or scheduler.
In practical vehicular systems, such integration can increase validation and maintenance cost.
Therefore, it is valuable to clarify how much benefit can be obtained from a simpler add-on design that preserves the existing MPQUIC scheduler behavior.

Motivated by this gap, we study an adaptive-FEC design for MPQUIC Datagram.
Rather than redesigning or assuming a specific MPQUIC scheduler, the proposed method treats the scheduler as an interchangeable component: it preserves the scheduler's original path-selection logic and applies sender-side adaptive redundancy control based on per-path loss estimates.
More specifically, it reuses the standard QUIC loss detection mechanism~\cite{rfc9002} to estimate smoothed per-path loss rates and determines the number of parity packets for each FEC block according to the expected packet loss.
The method is implemented using an existing MPQUIC implementation together with an off-the-shelf Reed--Solomon (RS) coding library, thereby improving practical deployability in existing systems.

We evaluate the proposed approach through vehicular field experiments over three commercial LTE/5G paths.
Our goal is not to claim the first adaptive-FEC mechanism for multipath QUIC, but to identify and validate a practical design point: a scheduler-agnostic adaptive-FEC controller that can be integrated into an existing MPQUIC stack and still provide measurable gains in real vehicular mobile environments~\cite{vu2021supporting,cellfusion2023}.
The experimental results show that, under the main \SI{4.0}{Mbps} setting, the proposed scheme reduces the average one-way delay from \SI{103.0}{ms} to \SI{70.8}{ms}, the 95th-percentile delay from \SI{281.2}{ms} to \SI{142.3}{ms}, and the packet loss rate from 1.7\% to 0.8\% compared with No-FEC.

The main contributions of this paper are as follows:
\begin{itemize}
\item We present a scheduler-agnostic adaptive-FEC controller for MPQUIC Datagram. The controller is scheduler-agnostic and operates on the packet allocation produced by the underlying MPQUIC scheduler, without modifying its path-selection logic.
\item We show a practical integration approach that reuses QUIC loss detection signals and an existing RS coding library, avoiding an additional path-quality probing subsystem or a specialized scheduler-coding interface.
\item Through vehicular field experiments over three commercial LTE/5G paths, we demonstrate that the proposed design improves both latency and reliability with low average redundancy under the measured mobile communication setting.
\end{itemize}

%======================
%======================
\section{Related Work}

\subsection{Multipath Transport}

Multipath transport has long been studied to improve throughput, robustness, and connection continuity over heterogeneous networks.
A representative example is MPTCP, which enables a single connection to simultaneously use multiple paths~\cite{rfc8684}.
MPTCP is effective in improving resilience against path degradation and in aggregating available path resources.
However, for delay-sensitive applications, MPTCP still inherits TCP's retransmission-based recovery and in-order delivery semantics, which can increase latency through retransmissions and HoL blocking.

To address these limitations, MPQUIC has been proposed as a QUIC-based multipath transport protocol~\cite{deconinck2017mpquic}.
QUIC itself is a UDP-based transport protocol that supports multiplexing, encryption, and user-space implementation, making it attractive for low-latency and rapidly evolvable transport services~\cite{rfc9000}.
In addition, the QUIC Datagram extension enables unreliable transmission without retransmissions, which is particularly useful for real-time applications that prefer timeliness over strict reliability~\cite{rfc9221}.
MPQUIC extends this framework to multiple paths, and its standardization is currently under discussion in the IETF~\cite{ietf_multipath_quic}.

A number of prior studies have also investigated scheduler designs for MPQUIC in order to improve performance under dynamic network conditions, including mobility-aware path selection~\cite{yang2022scheduler}.
These works show that path scheduling is an important factor in multipath transport performance.
However, they primarily focus on path selection and packet scheduling, and do not directly address how packet losses should be recovered in a low-latency manner when QUIC Datagram is used.

\subsection{Reliability over Multipath Communication}

Improving reliability over multipath communication has also been widely studied.
One simple approach is packet replication, in which duplicate packets are transmitted over multiple paths~\cite{markopoulou2007rail}.
Replication can reduce recovery delay and mitigate the impact of path-specific losses, and is therefore attractive for latency-sensitive communication~\cite{markopoulou2007rail}.
However, replication incurs a large bandwidth overhead because each protected packet is sent multiple times.
Its efficiency also degrades when multiple paths experience correlated impairments~\cite{zhang2010correlation}.

Coding-based approaches such as FEC provide a more bandwidth-efficient alternative.
Instead of duplicating each packet, FEC introduces coded redundancy across multiple packets so that a receiver can recover from some losses without waiting for retransmissions~\cite{rfc6363}.
This property makes FEC particularly suitable for real-time traffic.

\subsection{FEC for QUIC and Adaptive Redundancy Control}

Recent work has revisited FEC in the context of QUIC.
In particular, QUIC-FEC studies have shown that adding coded redundancy at the QUIC layer can improve robustness in lossy or high-delay environments~\cite{michel2019quicfec}.
These studies establish that QUIC can benefit from coding-based recovery, but they mainly focus on QUIC-level loss recovery itself and do not specifically target practical vehicular multipath deployments.

Adaptive-FEC has also been investigated to avoid the inefficiency of fixed redundancy.
In such approaches, the amount of parity is adjusted according to observed network conditions such as packet loss rate or delay variation.
For MPQUIC in particular, Vu and Wolff studied delay-sensitive applications with forward erasure correction and proposed both an adaptive-FEC mechanism and an FEC-aware scheduler~\cite{vu2021supporting}.
Their results demonstrated that combining adaptive coding with multipath transport can significantly reduce application-level delay in dynamic emulated environments.

More recently, CellFusion demonstrated the practical benefit of combining coded multipath transport with vehicle-to-cloud real-time video streaming in real deployments~\cite{cellfusion2023}, showing that coded multipath communication is useful in realistic vehicular environments. However, CellFusion is not a direct comparison target for this paper because it studies a tightly integrated design for high-quality vehicular video streaming, whereas we focus on a scheduler-agnostic adaptive-FEC controller that can be added to an existing MPQUIC Datagram stack.

\subsection{Gap and Positioning of This Work}

From the above literature, three observations can be made.
First, multipath transport is already a well-established means of improving robustness, and MPQUIC provides a promising low-latency platform for multipath communication.
Second, coding-based recovery and adaptive redundancy control are already known to be effective, including in MPQUIC-based settings.
Third, recent work such as CellFusion has shown that coded multipath communication can be useful in real vehicular scenarios.

However, two gaps remain for practical deployment in our target scenario.
One is an evaluation gap: the effectiveness of adaptive-FEC over MPQUIC Datagram has not yet been sufficiently characterized through field experiments over commercial vehicular LTE/5G paths.
The other is a deployability gap: it is still unclear how much performance improvement can be obtained from a scheduler-agnostic adaptive-FEC design.
This trade-off matters in practice because the cost of integrating a new mechanism into an existing MPQUIC implementation is often as important as the achievable performance gain.

%======================
\section{System Model and Problem Definition}

\subsection{System Model}

\begin{figure}[t]
  \centering
  \includegraphics[width=0.9\linewidth]{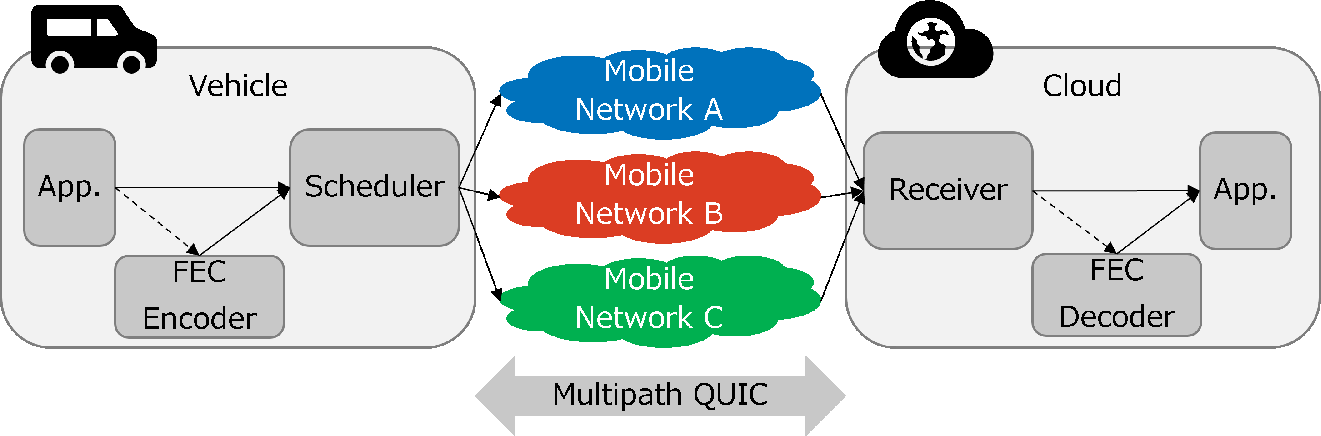}
  \caption{System overview of the proposed adaptive-FEC scheme over MPQUIC Datagram. Application data is encoded into FEC blocks at the sender, transmitted over multiple cellular paths using the underlying MPQUIC scheduler, and decoded at the receiver to recover packet losses.}
  \label{fig:system}
\end{figure}

We consider a vehicular communication system in which a vehicle transmits application data to a remote server over multiple cellular paths using MPQUIC~\cite{deconinck2017mpquic, ietf_multipath_quic}.
The QUIC Datagram extension is used to support low-latency transmission without retransmissions~\cite{rfc9221}.

Fig.~\ref{fig:system} illustrates the proposed architecture.
The sender groups application data into FEC blocks, generates parity packets, and passes both data and parity packets to the underlying MPQUIC scheduler.
The receiver buffers packets on a per-block basis and decodes missing packets before delivery to the application.

Let $K$ denote the number of available paths.
A packet scheduler distributes outgoing data packets across these paths.
Let $k_i$ denote the number of data packets in an FEC block that are actually scheduled onto path $i$, where
\begin{equation}
  k = \sum_{i=1}^{K} k_i,
\end{equation}
and $k$ is the total number of data packets in the FEC block.
In the proposed design, $k_i$ is counted after the data packets in the FEC block have been passed to the MPQUIC scheduler and their transmission paths have been determined.
Thus, $k_i$ is an observed scheduler outcome rather than a control variable of the FEC mechanism.
This assumption reflects our design objective of preserving the scheduler behavior and introducing adaptive-FEC as an add-on mechanism rather than as a cross-layer scheduler redesign.

\subsection{FEC Model}

We consider block-based FEC, where each block consists of $k$ data packets and $r$ parity packets.
The coding scheme allows recovery from up to $r$ packet losses within a block~\cite{rfc6363}.
We focus on block-level redundancy control and do not redesign the underlying MPQUIC packet scheduler or derive a path-dependent coding structure.

\subsection{Expected Packet Loss}

Given the data-packet allocation $k_i$ and per-path loss rates $p_i$, the expected number of lost data packets in one FEC block is approximated as
\begin{equation}
  \hat{L} = \sum_{i=1}^{K} k_i p_i.
\end{equation}

Here, $\hat{L}$ is computed only from the data packets in the FEC block.
The path allocation of parity packets is not included in this estimate because parity packets are generated after the number of redundant packets has been determined.
Instead, the proposed controller uses a safety factor in the redundancy decision to compensate for the mismatch between the data-packet-based expected loss and the actual packet losses observed over both data and parity transmissions.

This controller does not explicitly optimize the block recovery probability under a binomial or burst-loss model.
Instead, it intentionally uses the expected loss as a lightweight control signal in order to preserve scheduler independence and implementation simplicity.

\subsection{Problem Definition}

Our objective is to determine the number of parity packets $r$ for each FEC block so that redundancy tracks the expected block loss under time-varying per-path conditions while keeping overhead small.
More specifically, we seek a sender-side redundancy controller that improves robustness for MPQUIC Datagram without sacrificing the practical deployability of existing implementations.

The controller must avoid retransmission-based recovery, introduce redundancy only when necessary, adapt to time-varying per-path losses, and work with an existing MPQUIC implementation without assuming a scheduler-specific path-selection rule.
Accordingly, we formulate the problem as adaptive redundancy control based on per-path loss estimation and scheduler-determined packet allocation, rather than as joint optimization of scheduling and coding.

%======================
\section{Proposed Method}

\subsection{Design Overview}

The proposed method performs sender-side adaptive-FEC control for MPQUIC Datagram based on per-path loss estimation.
Its main design objective is to provide scheduler-agnostic loss protection, improving loss robustness without redesigning multipath scheduling or altering the existing MPQUIC scheduler behavior.
Accordingly, the method is \emph{scheduler-agnostic}: it observes the packet allocation produced by the scheduler, but does not modify the scheduler's path-selection logic.

The method reuses three existing components: the MPQUIC scheduler, the standard QUIC loss detection mechanism, and an off-the-shelf RS coding library.

The overall control flow is as follows.
The sender first updates per-path loss estimates using QUIC loss detection signals.
It then computes the expected loss of each FEC block from the scheduler-determined packet allocation and the current per-path loss estimates.
Based on this expected loss, the sender determines the number of parity packets, applies RS encoding, and transmits both data and parity packets over multiple paths.

\subsection{Block-Based Encoding}

For each FEC block, the sender groups $k$ QUIC Datagram payloads and generates $r$ parity packets using RS coding.
Each QUIC Datagram payload is treated as one RS source symbol.
In other words, FEC is applied to application data carried in QUIC Datagram frames, rather than to the entire QUIC packet including transport headers.

This block-based design is intentionally simple.
It avoids transport-header-level redesign and allows the coding function to be added as an application-data protection layer on top of an existing MPQUIC implementation.

\subsection{Per-Path Loss Estimation}

The sender-side smoothed loss-rate estimator follows the event-driven smoothed loss-rate formulation in~\cite{park2012tcp}.
In~\cite{park2012tcp}, the loss rate is estimated from the packet transmission interval between consecutive TCP loss events and then smoothed over time.
We use this smoothed loss-rate estimator as the basis for per-path loss estimation.

In our MPQUIC implementation, the estimator is applied on a per-path basis by replacing TCP loss events with QUIC loss detection events defined in RFC 9002~\cite{rfc9002}.
For each path, the sender counts the packets transmitted since the previous loss detection event on the same path and the packets newly declared lost at the current event.
The resulting loss observation is then incorporated into the smoothed loss-rate estimate following the method in~\cite{park2012tcp}.

If no new loss detection event occurs on a path, the previous smoothed loss-rate estimate for that path is retained until the next event.
Thus, the estimator remains event-driven, while its triggering and packet-counting logic are adapted to QUIC's per-path loss detection mechanism.
By reusing existing QUIC loss signals, the proposed method avoids introducing a separate path-quality estimation subsystem.

\subsection{Adaptive Redundancy Control}

Using the expected loss $\hat{L}$ defined in the previous section, the number of parity packets is determined as
\begin{equation}
  r = \min \left( k, \left\lceil \alpha \hat{L} \right\rceil \right),
\end{equation}
where $\alpha \ge 1$ is a safety factor.

The role of $\alpha$ is to compensate for the mismatch between the expected data-packet loss and the actual loss observed within each FEC block.
Although $\hat{L}$ reflects the average loss tendency of the paths used by the data packets, instantaneous variations, burst losses, and losses of parity packets are not explicitly modeled in the expected-loss calculation.
A safety margin is therefore introduced to improve the probability of successful recovery while keeping the controller simple and scheduler-agnostic.

In this study, we set $\alpha = 2.0$ as a conservative heuristic based on preliminary trace-level observations.
This value was chosen to reduce under-protection during temporary loss increases while keeping average redundancy low.
We do not claim that this value is universally optimal; sensitivity analysis of $\alpha$ is left for future work.

The value of $r$ is bounded between $0$ and $k$ to avoid excessive redundancy.
As a result, the controller increases parity when the estimated loss level is high and reduces parity when the network condition is favorable.

\subsection{Transmission and Reception}

Both data and parity packets are transmitted using the underlying MPQUIC scheduler.
The proposed method does not explicitly assign packets to paths beyond the scheduler's original behavior.
At the receiver, arriving packets are buffered on a per-block basis.
When at least $k$ packets belonging to the same block are available, RS decoding is performed and recovered data packets are delivered to the application.

If fewer than $k$ packets are available for a block, decoding cannot recover all missing data packets.
A block is considered incomplete when the receiver-side buffering timeout or block-closing condition is reached before $k$ packets are collected.
In this case, the receiver does not discard the entire FEC block.
Instead, the data packets that have already arrived are delivered to the application, while only unrecovered missing data packets are counted as lost.
This behavior is important for real-time Datagram traffic because it avoids converting partial block loss into whole-block loss.

\subsection{Implementation Considerations}

The implementation is built on our prior MPQUIC implementation~\cite{iwasawa} using the open-source quiche stack as the underlying QUIC software base\footnote{https://github.com/cloudflare/quiche}. It reuses existing QUIC loss detection and MPQUIC scheduling, and adds only sender-side and receiver-side RS encoding and decoding functions\footnote{https://github.com/AndersTrier/reed-solomon-simd} without requiring a specialized scheduler–coding interface.

%======================
\section{Field Experiment}

\subsection{Experimental Setup}

To evaluate the practical effectiveness of the proposed method in realistic mobile environments, we conducted vehicular field experiments using a vehicle equipped with multiple commercial LTE/5G cellular connections.
The experiments are intended to emulate a vehicle-to-cloud remote monitoring scenario, in which application data is continuously transmitted from a vehicle to a cloud-side receiver.

The sender runs on an in-vehicle PC (EAC5100) with Ubuntu 22.04, and the receiver is deployed in a cloud environment.
The sender transmits UDP traffic using \texttt{iperf2} (v2.2.1)\footnote{\url{https://sourceforge.net/projects/iperf2/}} at a constant offered load of \SI{4.0}{Mbps}.
We use this as the main evaluation setting because it represents a moderate-load operating point for continuous uplink transmission in remote monitoring scenarios.
In remote driving and teleoperation systems, real-time video streams constitute a major portion of uplink traffic and typically require sustained Mbps-level bandwidth to maintain situational awareness~\cite{3gpp_ts_22186,denouden2022remote}.
Thus, \SI{4.0}{Mbps} was selected as a representative condition for evaluating whether the proposed method can improve both latency and reliability under practical vehicular communication load.

The system uses three LTE/5G paths from two different carriers.
The vehicle travels along a fixed route in Musashino City, Tokyo, Japan, with one lap taking approximately 15 minutes.
The maximum vehicle speed is approximately \SI{60}{km/h}, representing typical urban driving conditions.
For the FEC configuration, the block size is set to $k = 20$ packets.
This value is chosen based on the application delay requirement and the packet transmission rate.

\subsection{Evaluation Metrics}

We evaluate the proposed method using packet loss rate (PLR), one-way delay (OWD), 95th-percentile delay ($p95$), and coding rate.
These metrics are selected to reflect the primary requirement of the target scenario, namely, low-latency and reliable delivery for remote monitoring~\cite{3gpp_ts_22186,denouden2022remote}.

In this paper, the coding rate is defined as $k/(k+r)$, where $k$ and $r$ denote the numbers of data and parity packets in each FEC block, respectively.
This metric is used to quantify redundancy overhead: a lower coding rate indicates that more parity packets are introduced.

The OWD values are obtained from the delay statistics output by the \texttt{iperf2} server.
Therefore, the measured OWD represents the application-level delay observed at the receiver side.
For the proposed FEC scheme, this measurement includes the buffering and decoding delay introduced before packets are delivered to the application.
The sender-side timestamp embedded in each original data packet is preserved through FEC recovery, and the delay is calculated when the packet is delivered to the \texttt{iperf2} server.
Accordingly, the reported delay values reflect the end-to-end effect of FEC recovery rather than only the network propagation and queuing delay.

For OWD measurement, the clocks of the in-vehicle sender and the cloud-side receiver are synchronized using NTP with chrony.\footnote{\url{https://chrony-project.org/}}
Although this method may include residual clock synchronization error, the same procedure is consistently applied to all compared cases.
In our measurement environment, the residual clock error is on the order of a few milliseconds, which is sufficiently smaller than the OWD differences discussed in this paper.

Among these metrics, the average OWD reflects typical delay performance, the 95th-percentile OWD captures delay tail behavior, and PLR captures residual unrecovered packet loss.
Since tail delay and packet loss directly affect the usability and safety of remote monitoring systems, these metrics are particularly important in assessing the practical value of the proposed method.

%======================
\section{Results and Discussion}

\begin{table}[t]
  \centering
  \footnotesize
  \setlength{\tabcolsep}{4pt}
  \caption{Performance comparison between the proposed adaptive-FEC scheme and No-FEC, aggregated over seven laps of the field experiment under the same \SI{4.0}{Mbps} offered load.}
  \label{tab:performance_main}
  \begin{tabular}{lcccc}
    \hline
    Method & Avg. OWD [ms] & p95 OWD [ms] & PLR [\%] & Coding Rate \\
    \hline
    No-FEC   & 103.0 & 281.2 & 1.7 & 1.00 \\
    Proposed &  70.8 & 142.3 & 0.8 & 0.94 \\
    \hline
  \end{tabular}
\end{table}

\begin{figure}[t]
  \centering
  \includegraphics[width=\columnwidth]{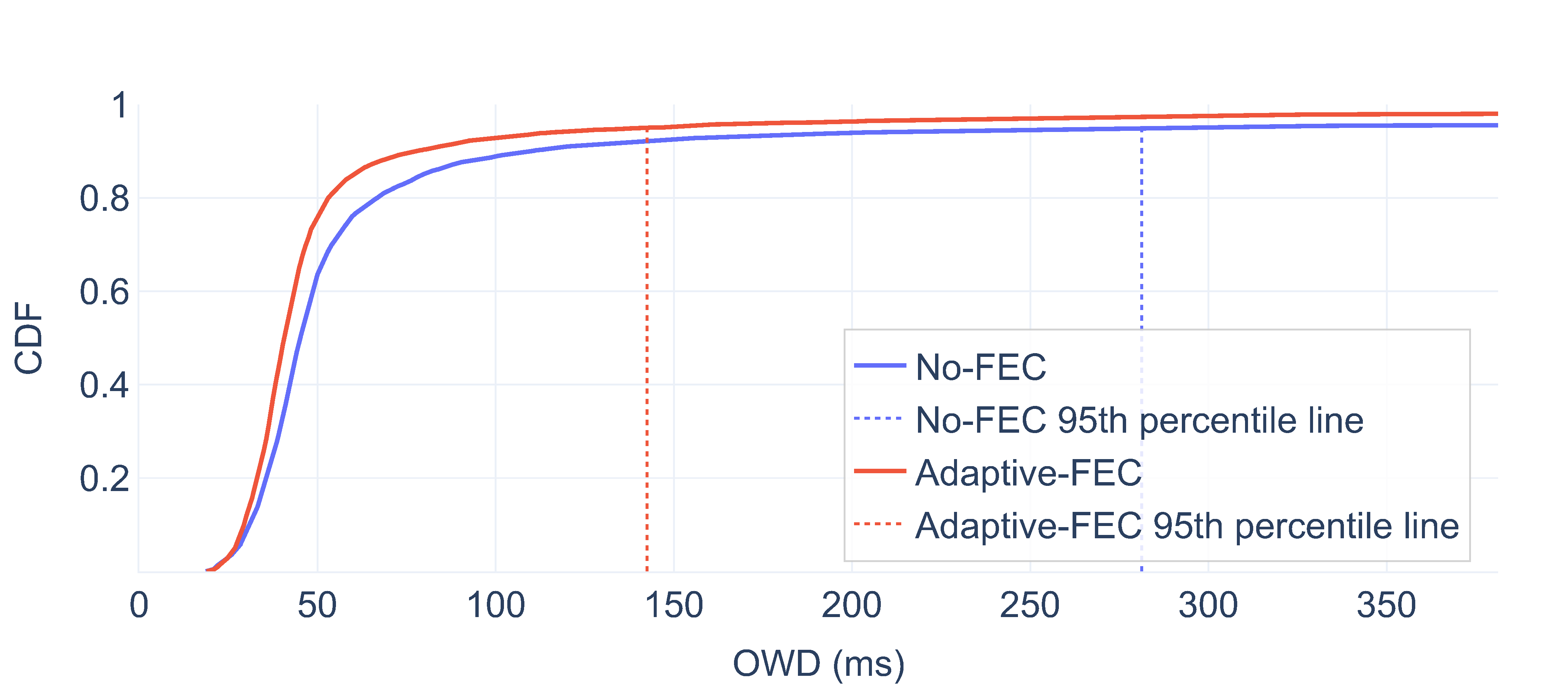}
  \caption{Empirical CDF of packet-level one-way delay (OWD) pooled over seven laps under the same \SI{4.0}{Mbps} offered load. The proposed adaptive-FEC scheme shifts the distribution toward lower delay values and reduces the 95th-percentile delay compared with No-FEC. The x-axis is truncated for readability.}
  \label{fig:owd_cdf}
\end{figure}

Table~\ref{tab:performance_main} presents the main comparison between the proposed method and No-FEC under the same \SI{4.0}{Mbps} offered load, aggregated over seven laps of the field experiment.

In the main \SI{4.0}{Mbps} setting, the proposed adaptive-FEC scheme improves both reliability and latency relative to No-FEC.
The average PLR decreases from 1.7\% to 0.8\%, the average OWD from \SI{103.0}{ms} to \SI{70.8}{ms}, and the 95th-percentile OWD from \SI{281.2}{ms} to \SI{142.3}{ms}.
The average coding rate is approximately 0.94, indicating that these gains are obtained with low average redundancy overhead.

Although FEC introduces additional buffering and decoding overhead, it can also reduce application-level delay in multipath communication.
When packets in the same FEC block are transmitted over different paths, some parity packets may arrive earlier than delayed original data packets.
Once the receiver obtains any $k$ packets among the $k$ data packets and $r$ parity packets in the block, RS decoding can recover the missing data packets without waiting for the delayed original packets.
Thus, the proposed scheme can mask not only packet losses but also path-specific delay spikes.
This behavior explains why the measured OWD improves despite the use of FEC, especially under vehicular cellular conditions where individual paths can temporarily experience delay spikes.

Fig.~\ref{fig:owd_cdf} shows the empirical CDF of packet-level OWD pooled over the seven laps under the same \SI{4.0}{Mbps} offered load.
The proposed method shifts the OWD distribution toward lower delay values over most of the range and also reduces the 95th-percentile delay compared with No-FEC.
This result is important for remote monitoring applications because both typical delay and tail delay affect practical responsiveness.

We further examined path diversity by computing Pearson correlations between pairs of per-path smoothed loss-rate series averaged over aligned 1-s bins.
Across the three paths, the total number of per-path 1-s bins was 19,675.
The pairwise correlations were weak (0.082, -0.052, and 0.113), indicating limited linear dependence among the three LTE/5G paths.

Overall, the field results suggest that adaptive-FEC over MPQUIC Datagram can improve both reliability and latency in vehicular mobile environments with moderate average redundancy overhead.
At the same time, the present results should be interpreted within the scope of this study, which focuses on a scheduler-agnostic implementation and a limited set of offered-load conditions rather than an exhaustive comparison across all redundancy baselines.

%======================
\section{Conclusion}

This paper presented an adaptive-FEC controller for MPQUIC Datagram that reuses QUIC loss detection signals and operates in a scheduler-agnostic manner.
Vehicular field experiments over three commercial LTE/5G paths showed that, under the main \SI{4.0}{Mbps} offered load, the proposed method reduced both latency and packet loss relative to No-FEC while maintaining a high average coding rate of 0.94.
In particular, the average OWD was reduced from \SI{103.0}{ms} to \SI{70.8}{ms}, the 95th-percentile OWD from \SI{281.2}{ms} to \SI{142.3}{ms}, and the PLR from 1.7\% to 0.8\%.
We also observed weak cross-path loss correlation among the three LTE/5G paths, which supports the use of redundancy over multipath transmission in the measured vehicular environment.

The main practical takeaway is that measurable gains in both reliability and latency can already be obtained from a scheduler-agnostic adaptive-FEC design, even without redesigning the MPQUIC scheduler.
Thus, the contribution of this work is not to claim the first coded or adaptive MPQUIC system, but to validate a practical design point for deployable MPQUIC Datagram enhancement in real vehicular mobile environments.

Future work includes controlled comparison against matched-overhead baseline schemes, sensitivity analysis of safety factor $\alpha$ and FEC block size $k$, and evaluation under a wider range of offered loads, traffic types, and mobile environments.

\bibliographystyle{IEEEtran}
\bibliography{references}

@misc{rfc8684,
  author       = {Alan Ford and Costin Raiciu and Mark Handley and Olivier Bonaventure and Christoph Paasch},
  title        = {{TCP} Extensions for Multipath Operation with Multiple Addresses},
  howpublished = {RFC 8684},
  month        = mar,
  year         = {2020},
  doi          = {10.17487/RFC8684}
}

@misc{rfc9000,
  author       = {Jana Iyengar and Martin Thomson},
  title        = {{QUIC}: A {UDP}-Based Multiplexed and Secure Transport},
  howpublished = {RFC 9000},
  month        = may,
  year         = {2021},
  doi          = {10.17487/RFC9000}
}

@misc{rfc9002,
  author       = {Jana Iyengar and Ian Swett},
  title        = {{QUIC} Loss Detection and Congestion Control},
  howpublished = {RFC 9002},
  month        = may,
  year         = {2021},
  doi          = {10.17487/RFC9002}
}

@misc{rfc9221,
  author       = {Tommy Pauly and Eric Kinnear and David Schinazi},
  title        = {An Unreliable Datagram Extension to {QUIC}},
  howpublished = {RFC 9221},
  month        = mar,
  year         = {2022},
  doi          = {10.17487/RFC9221}
}

@misc{rfc6363,
  author       = {Watson, M. and Begen, A. and Roca, V.},
  title        = {Forward Error Correction ({FEC}) Framework},
  howpublished = {RFC 6363},
  month        = oct,
  year         = {2011},
  doi          = {10.17487/RFC6363}
}

@inproceedings{deconinck2017mpquic,
  author    = {De Coninck, Quentin and Olivier Bonaventure},
  title     = {{Multipath QUIC}: Design and Evaluation},
  booktitle = {Proc. 13th International Conference on Emerging Networking EXperiments and Technologies ({CoNEXT})},
  address   = {Incheon, Republic of Korea},
  pages     = {160--166},
  month     = dec,
  year      = {2017},
  doi       = {10.1145/3143361.3143370}
}

@misc{ietf_multipath_quic,
  author       = {Yanmei Liu and Yunfei Ma and De Coninck, Quentin and Olivier Bonaventure and Christian Huitema and Mirja K{\"u}hlewind},
  title        = {Managing Multiple Paths for a {QUIC} Connection},
  howpublished = {{IETF} Internet-Draft, draft-ietf-quic-multipath-21},
  year         = {2026},
  month        = mar,
  note         = {Work in progress},
  url          = {https://datatracker.ietf.org/doc/draft-ietf-quic-multipath/21/}
}

@inproceedings{yang2022scheduler,
  author    = {Wenjun Yang and Lin Cai and Shengjie Shu and Jianping Pan},
  title     = {Scheduler Design for Mobility-Aware Multipath {QUIC}},
  booktitle = {Proc. 2022 IEEE Global Communications Conference (GLOBECOM)},
  month     = dec,
  year      = {2022},
  address   = {Rio de Janeiro, Brazil},
  pages     = {2849--2854},
  doi       = {10.1109/GLOBECOM48099.2022.10001247}
}

@inproceedings{michel2019quicfec,
  author    = {Fran{\c{c}}ois Michel and De Coninck, Quentin and Olivier Bonaventure},
  title     = {{QUIC-FEC}: Bringing the Benefits of Forward Erasure Correction to {QUIC}},
  booktitle = {Proc. 2019 IFIP Networking Conference (IFIP Networking)},
  pages     = {1--9},
  address   = {Warsaw, Poland},
  month     = may,
  year      = {2019},
  doi       = {10.23919/IFIPNetworking.2019.8816838}
}

@inproceedings{vu2021supporting,
  author    = {Vu Anh Vu and Jan Wolff},
  title     = {Supporting Delay-Sensitive Applications with Multipath {QUIC} and Forward Erasure Correction},
  booktitle = {Proc. 17th ACM Symposium on QoS and Security for Wireless and Mobile Networks (Q2SWinet '21)},
  address   = {Alicante, Spain},
  month     = nov,
  year      = {2021},
  pages     = {95--103},
  doi       = {10.1145/3479242.3487312}
}

@techreport{3gpp_ts_22186,
  author      = {{3rd Generation Partnership Project (3GPP)}},
  title       = {Service Requirements for Enhanced {V2X} Scenarios},
  institution = {{3GPP}},
  type        = {Technical Specification},
  number      = {{TS} 22.186, ver. 19.0.0},
  month       = oct,
  year        = {2025}
}

@article{denouden2022remote,
  author  = {Jos den Ouden and Victor Ho and Tijs van der Smagt and Geerd Kakes and Simon Rommel and Igor Passchier and Jakub Juza and Idelfonso Tafur Monroy},
  title   = {Design and Evaluation of Remote Driving Architecture on {4G} and {5G} Mobile Networks},
  journal = {Frontiers in Future Transportation},
  volume  = {2},
  month   = jan,
  note    = {Art. no. 801567},
  year    = {2022},
  doi     = {10.3389/ffutr.2021.801567}
}

@inproceedings{cellfusion2023,
  author    = {Yunzhe Ni and Zhilong Zheng and Xianshang Lin and Fengyu Gao and Xuan Zeng and Yirui Liu and Tao Xu and Hua Wang and Zhidong Zhang and Senlang Du and Guang Yang and Yuanchao Su and Dennis Cai and Hongqiang Harry Liu and Chenren Xu and Ennan Zhai and Yunfei Ma},
  title     = {{CellFusion}: Multipath Vehicle-to-Cloud Video Streaming with Network Coding in the Wild},
  booktitle = {{Proc. ACM SIGCOMM 2023 Conference (SIGCOMM '23)}},
  address   = {New York, NY, USA},
  month     = sep,
  pages     = {668--683},
  year      = {2023},
  doi       = {10.1145/3603269.3604832}
}

@article{markopoulou2007rail,
  author  = {Athina Markopoulou and David R. Cheriton},
  title   = {The Case for Redundant Arrays of Internet Links ({RAIL})},
  journal = {Computing Research Repository (CoRR)},
  volume  = {abs/cs/0701133},
  year    = {2007}
}

@inproceedings{zhang2010correlation,
  author    = {Xin Zhang and Adrian Perrig},
  title     = {Correlation-Resilient Path Selection in Multi-Path Routing},
  booktitle = {{Proc. 2010 IEEE Global Telecommunications Conference (GLOBECOM 2010)}},
  address   = {Miami, FL, USA},
  month     = dec,
  year      = {2010},
  pages     = {2140--2145},
  doi       = {10.1109/GLOCOM.2010.5683582}
}

@inproceedings{iwasawa,
  title        = {Time in Flight Based {MPQUIC} Scheduler for Remote Monitoring of Autonomous Driving},
  author       = {Hiroki Iwasawa and Taichi Kawano and Seiya Komatsu and Takuya Tojo and Takeshi Kuwahara},
  booktitle    = {Proc. 2025 IEEE 22nd Consumer Communications \& Networking Conference (CCNC)},
  month        = jan,
  year         = {2025},
  pages        = {1--4},
  doi          = {10.1109/ccnc54725.2025.10975968},
  address      = {Las Vegas, NV, USA},
}

@article{park2012tcp,
  author  = {M.-Y. Park and S.-H. Chung and C.-W. Ahn},
  title   = {{TCP}'s dynamic adjustment of transmission rate to packet losses in wireless networks},
  journal = {{EURASIP Journal on Wireless Communications and Networking (JWCN)}},
  note    = {Art. no. 304},
  volume  = {2012},
  year    = {2012},
  month   = sep,
  doi     = {10.1186/1687-1499-2012-304}
}

\end{document}